\overfullrule=0pt
\input harvmac
\def\a{{\alpha}}
\def\ah{{\overline\alpha}}

\def\l{{\lambda}}

\def\kh{{\overline\kappa}}
\def\b{{\beta}}
\def\bh{{\overline\beta}}
\def\g{{\gamma}}
\def\gh{{\overline\gamma}}
\def\d{{\delta}}
\def\dh{{\overline\delta}}
\def\e{{\epsilon}}
\def\s{{\sigma}}

\def\half{{1\over 2}}
\def\p{{\partial}}

\def\t{{\theta}}

\Title{ \vbox{\baselineskip12pt
\hbox{IFT-P.020/2002}}}
{\vbox{\centerline
{Conformal Field Theory for the Superstring in a}
\centerline{Ramond-Ramond Plane Wave Background}}}
\smallskip
\centerline{Nathan Berkovits\foot{e-mail: nberkovi@ift.unesp.br}}
\smallskip
\centerline{\it 
Instituto de F\'{\i}sica Te\'orica, Universidade Estadual Paulista}
\centerline{\it
Rua Pamplona 145, 01405-900, S\~ao Paulo, SP, Brasil}
\bigskip

\noindent

A quantizable worldsheet action is constructed for the 
superstring in a supersymmetric plane wave background with Ramond-Ramond flux.
The action is manifestly
invariant under all isometries of the background and is an exact
worldsheet conformal field theory.

\Date{March 2002}

\newsec{Introduction}

It has recently been recognized that the Penrose limit of the
$AdS_5\times S^5$ background with Ramond-Ramond (R-R) flux is
a supersymmetric plane wave \ref\ppw{M. Blau,
J. Figueroa-O'Farrill, C. Hull and G. Papadopoulos,
{\it A New Maximally Supersymmetric Background of
IIB Superstring Theory}, JHEP 0201 (2002) 047,
hep-th/0110242.}, and the superstring
in this background is described in the light-cone
Green-Schwarz (GS) formalism \ref\gslc{M.B. Green and
J.H. Schwarz, {\it Supersymmetrical Dual String Theory},
Nucl. Phys. B181 (1981) 502.} by a quadratic worldsheet action
\ref\metsaev{R.R. Metsaev, {\it Type IIB Green-Schwarz Superstring
in Plane Wave Ramond-Ramond Background}, Nucl. Phys. B625 (2002) 70,
hep-th/0112044.}
\ref\russo{J.G. Russo and A.A. Tseytlin, {\it On Solvable Models of
Type IIB Superstring in NS-NS and R-R Plane Wave Backgrounds},
hep-th/0202179.}.
The spectrum of physical states with nonzero $P^+$ momentum 
can easily be computed using this light-cone GS action, which has
been useful for checking aspects of the Maldacena conjecture
\ref\mald{D. Berenstein, J. Maldacena and H. Nastase, {\it
Strings in Flat Space and PP Waves from N=4 Super Yang-Mills},
hep-th/0202021.}.
However, to compute tree amplitudes or to describe physical states
with vanishing $P^+$ momentum, the light-cone GS formalism is
problematic even in a flat background. As will be discussed below,
the problems associated with light-cone gauge become even more
troublesome in the plane wave background since there is no $J^{+-}$
isometry.
Furthermore, it would be convenient to have a quantizable worldsheet
action in which all isometries of the plane wave background are manifest,
and not just those isometries which commute with the light-cone gauge choice.

Although the covariant GS action \ref\gscov{M.B. Green
and J.H. Schwarz, {\it Covariant Description of Superstrings},
Phys. Let. B136 (1984) 367.}\metsaev\ can classically describe 
the plane wave background in a manner which preserves all isometries,
it is not known how to covariantly quantize the GS action.
Over the last eight years, an alternative formalism for the superstring
has been developed which can be covariantly quantized \ref\berkfour
{N. Berkovits, {\it Covariant Quantization of the Green-Schwarz
Superstring in a Calabi-Yau Background}, Nucl. Phys. B431 (1994) 258,
hep-th/9404162.}\ref\topo{N. Berkovits and C. Vafa, {\it N=4
Topological Strings}, Nucl. Phys. B433 (1995) 123, hep-th/9407190.} 
\ref\berkten {N. Berkovits, {\it Super-Poincar\'e
Covariant Quantization of the Superstring}, JHEP 04 (2000) 018,
hep-th/0001035.}. 
In a flat
background, this formalism has a quadratic
worldsheet action 
and tree amplitudes can be easily computed in a manifestly super-Poincar\'e
covariant manner. The formalism generalizes to curved backgrounds \ref\siegberk
{N. Berkovits and W. Siegel, {\it Superspace Effective Actions for
4D Compactifications of Heterotic and Type II Superstrings}, Nucl. Phys.
B462 (1996) 213, hep-th/9510106.}
and was used to construct quantizable actions for the superstring in
$AdS_5\times S^5$ \berkten\ref\chandia{N. Berkovits and O. Chandia,
{\it Superstring Vertex Operators in an $AdS_5\times S^5$ Background},
Nucl. Phys. B596 (2001), hep-th/0009168.}, $AdS_3\times S^3$
\ref\adswitten
{N. Berkovits, C. Vafa and E. Witten, {\it Conformal Field Theory
of $AdS$ Background with Ramond-Ramond Flux}, JHEP 9903 (1999)
018, hep-th/9902098.}
\ref\six{N. Berkovits, {\it Quantization of the
Type II Superstring in a Curved Six-Dimensional Background},
Nucl. Phys. B565 (2000) 333, hep-th/9908041.}, 
and $AdS_2\times S^2$ 
\ref\bz
{N. Berkovits, M. Bershadsky, T. Hauer, S. Zhukov and B. Zwiebach,
{\it Superstring Theory on $AdS_2\times S^2$ as a Coset Supermanifold},
Nucl. Phys. B567 (2000) 61, hep-th/9907200.}
backgrounds
with R-R flux. 

In this paper, this alternative formalism will be used to construct
a quantizable action for the superstring in the
plane wave background obtained by taking the Penrose limit
of these $AdS_{D\over 2}
\times S^{D\over 2}$ backgrounds. By using the ``pure spinor'' or
``hybrid'' versions of the formalism, all isometries
of the plane wave background can be made manifest. 
The action is not quadratic, which
is not surprising since the action for the
bosonic plane wave background
is not quadratic when written in conformal gauge.
However, the action is simpler than its $AdS_{D\over 2}\times S^{D\over 2}$ 
counterpart and can be 
proven to be an exact
conformal field theory.

In section 2 of this paper, the limitations of the light-cone GS
formalism will be discussed. In section 3, a covariantly quantizable action
will be constructed for the Penrose limit of the 
$AdS_5\times S^5$, $AdS_3\times S^3$, and $AdS_2\times S^2$ backgrounds
with R-R flux. And in section 4, the action will be proven to
be an exact conformal field theory.

\newsec{Limitations of the Light-Cone GS Formalism}

Since the light-cone action only depends on physical worldsheet
variables, the light-cone formalism for the bosonic string and
superstring is the most efficient way to compute the physical
spectrum at non-zero $P^+$ momentum. However, besides the lack
of manifest covariance, there are various other drawbacks of the
light-cone GS action
which are not present in the covariant action.\foot{This
section is based on several discussions with Michael Green.}

One drawback is that the light-cone gauge is only well-defined
when $P^+$ is nonzero, so the light-cone action cannot be used
to obtain the spectrum at vanishing $P^+$ momentum. Although
in a flat background, one can always rotate any state with nonzero
momentum to have nonzero $P^+$ momentum, this is not always possible in
backgrounds such as the plane wave background where  
$SO(9,1)$ covariance is absent. So the light-cone formalism in
a plane wave background may be unable to describe certain non-trivial
physical states.

Another drawback of the light-cone formalism which is especially
problematic for the
light-cone GS formalism is the explicit dependence on interaction points
in the computation of scattering amplitudes.
Recall that $N$-point tree amplitudes in light-cone gauge are
computed using the Mandelstam map\ref\mandmap{S. Mandelstam,
{\it Interacting String Picture of Dual Resonance Models},
Nucl. Phys. B64 (1973) 205.} 
\eqn\mandel{\rho(z)= \sum_{r=1}^N P_r^+ \log(z-z_r)}
where $\rho(z)$ maps the complex plane into the interacting string
diagram and $P_r^+$ is the $P^+$ momentum of the $r^{th}$ external
string. For the bosonic string in light-cone gauge, interactions are
described by a simple overlap integral, so scattering amplitudes 
can be easily computed by evaluating correlation functions of light-cone
vertex operators located at $z=z_r$ in the complex plane, which
get mapped to $\rho=\pm\infty$ in the string diagram.

However, for the GS superstring in light-cone gauge, interactions are
not simply overlap integrals but also include an explicit operator which
must be inserted at the interaction point \ref\interpoint
{M.B. Green and J.H. Schwarz, {\it Superstring Interactions},
Nucl. Phys. B218 (1983) 43.}\ref\mandinter{S. Mandelstam, {\it Interacting 
String Picture of the Fermionic String}, Prog. Theor. Phys. Suppl.
86 (1986) 163.}.\foot{For the RNS
superstring in light-cone gauge, one also needs to include an
interaction point operator when using the Mandelstam map of \mandel\
to describe the string diagram \ref\threes{S. Mandelstam,
{\it Interacting String Picture of the Neveu-Schwarz-Ramond
Model}, Nucl. Phys. B69
(1974) 77.}. However, if one instead describes
the string diagram using the map $\rho(z,\t)=
\sum_{r=1}^N P_r^+ \log(z-z_r-\t\t_r)$ where $(z,\t)$ parameterizes
a complex ``super-plane'', one can avoid interaction point operators
in the light-cone RNS formalism \ref\mesuper{N. Berkovits,
{\it Calculation of Scattering Amplitudes for the Neveu-Schwarz
Model using Supersheet Functional Integration}, Nucl. Phys. B276 (1986) 650.}.}
Using $SU(4)\times U(1)$ notation,
this interaction point operator is 
\eqn\interact{|\p x_L +\p x_{[\mu\nu]} 
S^\mu S^\nu +{1\over{24}}\p x_R \e_{\mu\nu\rho\sigma} 
S^\mu S^\nu S^\rho S^\sigma|^2}
where $\mu=1$ to 4 is an $SU(4)$ index,
$[S^\mu,S_\mu]$ and $[\overline S^\nu, \overline S_\nu]$
are the left and right-moving SO(8) spinors decomposed in terms of
a $[4_\half,\overline 4_{-\half}]$ representation of $SU(4)\times U(1)$, and
$[x_L,x_{[\mu\nu]},x_R]$ is the SO(8) vector
decomposed in terms of
a $[1_1,6_0,1_{-1}]$ representation of $SU(4)\times U(1)$.
For an $N$-point tree amplitude described by the map of
\mandel, the $N-2$ interaction point
operators are located at the points $z_\kappa$ where 
\eqn\lcinter{
{{\p \rho}\over{\p z}}|_{z=z_\kappa}= \sum_{r=1}^N {{P_r^+}\over
{z_\kappa-z_r}}=0.}
So scattering amplitudes are computed by evaluating correlation functions
which involve both light-cone vertex operators at $z=z_r$ 
and interaction point operators at $z=z_\kappa$.
Since expressing $z_\kappa$ in terms of $z_r$ requires finding the zeros
of a polynomial of degree $N-2$, the light-cone GS formalism has not
yet been used to 
compute tree amplitudes with more than four external strings.\foot{
One trick \ref\gstrick{M.B. Green and J.H. Schwarz,
{\it Supersymmetrical Dual String Theory, 2. Vertices and Trees},
Nucl. Phys. B198 (1982) 252.}
for computing light-cone GS amplitudes in a flat background
is to choose a Lorentz frame in which $P_r^+\to 0$ for $r=2$ to $N-1$.
In this ``short string'' limit, $z_\kappa\to z_r$ for $r=2$ to $N-1$
and the interaction point operator combines with the light-cone vertex
operator to give an operator which resembles the covariant vertex operator
in light-cone gauge. After computing the scattering amplitude in this
``short string'' limit, one can then use SO(9,1) covariance to derive
the amplitude for generic values of $P_r^+$. However, this trick
does not work in the plane wave background because of the absence of
SO(9,1) covariance.}
Furthermore, singularities occuring when interaction points collide imply
that one needs to include contact terms in the light cone 
interaction to remove these singularities \ref\greens{J. Greensite
and F.R. Klinkhamer, {\it Superstring Amplitudes and Contact Interactions},
Nucl. Phys. B304 (1988) 108.}. 
The precise form of these light-cone contact terms has not been
worked out. Note that in a covariant formalism, these problems associated
with light-cone interaction point operators are not present since one
can always
``smooth out'' the interaction point using a conformal transformation.

A third drawback of the light-cone formalism is that in backgrounds
which are not invariant under the $J^{+-}$ Lorentz transformation, the
light-cone action is complicated when written in the complex plane.
For example, in the supersymmetric plane wave background, the light-cone
GS action is \metsaev\
\eqn\lcgs{{\cal S}= \int d^2 \rho (\half\p_\rho x^j 
\overline\p_{\overline\rho} x^j
+ S^a \overline\p_{\overline\rho} S^a + \overline S^a \p_\rho\overline S^a
+ \half \mu^2 x^j x^j + 2\mu S^a \s^{1234}_{ab} \overline S^b)}
where $\rho$ parameterizes the interacting string diagram, $a=1$ to 8 is
an SO(8) spinor index,
and $F^{-1234}= F^{-5678}= \mu$ is the R-R flux.

Using the Mandelstam map of \mandel, the action in the complex plane is 
therefore
\eqn\lcgsz{{\cal S}= \int d^2 z (\half\p_z x^j \overline\p_{\overline z} x^j
+ S_z^a \overline\p_{\overline z} S_z^a + 
\overline S_{\overline z}^a \p_z\overline S_{\overline z}^a}
$$
+{{\mu^2}\over 2} |\sum_{r=1}^N {{P_r^+}\over{z-z_r}}|^2
 x^j x^j + 2 \mu 
|\sum_{r=1}^N {{P_r^+}\over{z-z_r}}| S_z^a \s^{1234}_{ab} 
\overline S_{\overline z}^b)$$
where $S_z^a = ({{\p\rho}\over{\p z}})^\half S^a$ and
$\overline S_{\overline z}^a 
= ({{\p\overline\rho}\over{\p \overline z}})^\half
\overline S^a$.\foot{It is 
interesting to note that the $\mu$ dependence of the action of \lcgsz\
drops out near the light-cone interaction points $z_\kappa$ satisfying
\lcinter. This suggests that the light-cone interaction point operator
in a plane wave background is the same as the light-cone interaction
point operator of \interact\ in a flat background. I would like
to thank Michael Green for discussions on this point.}
This means that the Green's function $G^{jk}(z,z')$ for
$\langle x^j(z) z^k(z')\rangle$ in the complex plane
must satisfy the complicated differential equation
\eqn\greend{(\p_z\overline\p_{\overline z} - \mu^2
|\sum_{r=1}^N {{P_r^+}\over{z-z_r}}|^2) G^{jk}(z,z')= \eta^{jk}\d^2(z-z').}
Finding a solution to \greend\ is probably no easier than computing
OPE's using a conformally invariant action which is not quadratic.

So although the quadratic action in the light-cone GS formalism
is extremely useful for computing the physical spectrum at nonzero
$P^+$ momentum, it is not convenient for describing physical states
with vanishing $P^+$ momentum or for computing tree-level scattering
amplitudes.

\newsec{Covariant Action for the Superstring in R-R Plane Wave Background}

In this section, a quantizable action will be constructed for
the supersymmetic plane wave background coming from the Penrose limit
of the $AdS_5\times S^5$, $AdS_3\times S^3$ and $AdS_2\times S^2$
backgrounds with R-R flux. Although the action will have features in 
common with the covariant GS action in these backgrounds, there are
some important differences which allow covariant quantization.

One difference is the presence of the worldsheet variables
$d_\a$ and $\overline d_\ah$ which play the role of conjugate momenta
to the left and right-moving $\t^\a$ and $\overline \t^\ah$ variables.\foot{
Even though $\t^\a$ and $\overline\t^\ah$ have the same chirality for
the Type IIB superstring, it will be convenient to distinguish the
spinor indices on these left and right-moving variables by using
barred or unbarred indices.} These conjugate momentum variables
break kappa symmetry, which is replaced by BRST invariance
in the $D=10$ version of the action and by N=2 worldsheet superconformal
invariance in the $D=6$ and $D=4$ versions of the action.
A second difference with the covariant GS formalism is the presence
of bosonic worldsheet ghost variables. In the $D=10$ version of the action,
these worldsheet ghosts transform as pure spinors under Lorentz
transformations, while in the $D=6$ and $D=4$ versions, they
are Lorentz scalars.

Since the alternative formalism for the superstring
can be defined in
any consistent supergravity background, 
it is straightforward to construct the worldsheet action in any specific
background.
In either the $AdS_{D\over 2}\times S^{D\over 2}$ background with R-R flux
or its
corresponding plane wave limit,
the worldsheet action is
\eqn\altact{{\cal S} = {\cal S}_{GS} +
\int d^2 z (d_\a \overline L^\a + \overline d_\ah L^\ah -{1\over 2}
 d_\a \overline d_\bh F^{\a\bh})
+{\cal S}_{comp} + {\cal S}_{ghost}}
where 
$F^{m_1 ... m_{D\over 2}}$ is the constant $D\over 2$-form self-dual
Ramond-Ramond flux and
$F^{\a\bh}=$
$ {1\over{({D\over 2})!}} F^{m_1 ... m_{D\over 2}} 
(\g_{m_1 ... m_{D\over 2}})^{\a\bh}$.

The first term 
${\cal S}_{GS}$ in \altact\ is the standard covariant GS action
\eqn\covgs{{\cal S}_{GS}=\int d^2 z [\half \eta_{mn} L^m \overline L^n +
\int dy \e^{IJK} (\g_{m\a\b}L^m_I L^\a_J L^\b_K +
\g_{m\ah\bh} L^m_I L^\ah_J L^\bh_K)]}
constructed
using the Metsaev-Tseytlin currents
\ref\metsaevtseytlin{R. Metsaev and A. Tseytlin, {\it Type
IIB Superstring Action in $AdS_5\times S^5$ Background},
Nucl. Phys. B533 (1998) 109, hep-th/9805028.}\metsaev\ 
\eqn\current{G^{-1} \p G = P_m L^m + Q_\a L^\a +  Q_\ah L^\ah
+ \half J_{mn} L^{mn},}
$$G^{-1} \overline\p G 
= P_m \overline L^m + Q_\a \overline L^\a +  Q_\ah \overline L^\ah
+ \half J_{mn} \overline L^{mn},$$
where 
$G(x^m,\t^\a, \overline\t^\ah)= \exp(x^m P_m +
\t^\a Q_\a + \overline\t^\ah  Q_\ah)$,
$[x^m, \t^\a,\overline\t^\ah]$ are
$N=2$ $D$-dimensional superspace variables\foot{In $D=6$, the action of
\altact\
uses the 
hybrid superstring formalism with eight $\t$'s and eight $\overline\t$'s.
As discussed in \six, the $D=6$ action of \adswitten\ using the hybrid
superstring formalism
with four $\t$'s and four $\overline\t$'s can be obtained from \altact\
by using the ``harmonic'' constraint
to gauge away $(\t^{\a 2},\overline\t^{\overline\a 2})$ and 
to replace $(d_{\a 2},\overline d_{\overline\a 2})$
with $(d_{\a 1} e^{-\rho-i\sigma},
\overline d_{\overline \a 1} e^{-\overline \rho-i\overline \sigma})$.}
with 
$m=0$ to $D-1$ and $[\a,\ah]=1$ to $(2D-4)$, 
the generators
$[P_m,Q_\a, Q_\ah, J_{mn}]$ form a super-Lie algebra with the
commutation relations
\eqn\commrel{[P^m,P^n]= \half
R^{mnpq} J_{pq}, \quad  \{Q_\a,Q_\b\}=2\g^m_{\a\b} P_m,
\quad
\{Q_\ah,Q_\bh\}=2\g^m_{\ah\bh} P_m,}
$$
[Q_\a, P^m] = \g^m_{\a\b} F^{\b\gh} Q_\gh,\quad
[Q_\ah, P^m] = -\g^m_{\ah\bh} F^{\g\bh} Q_\g,\quad
\{Q_\a, Q_\gh\}= \half J_{[mn]} \g^m_{\a\b} F^{\b\dh} \g^n_{\dh\gh},$$ 
$J_{mn}$ generate the usual Lorentz algebra,
$R^{mnpq}$ is the spacetime curvature which is related to $F^{\a\bh}$
by the identity
\eqn\curvid{R^{mnpq}(\g_{pq})_\a^\b = \g^m_{\a\g} F^{\g\dh} \g^n_{\dh\kh}
F^{\b\kh} -
\g^n_{\a\g} F^{\g\dh} \g^m_{\dh\kh}
F^{\b\kh},}
$\g^m_{\a\b}$ and $\g^m_{\ah\bh}$ are $(2D-4)\times (2D-4)$ 
symmetric $\g$-matrices, and 
$\int dy \e^{IJK} (\g_{m\a\b}L^m_I L^\a_J L^\b_K +
\g_{m\ah\bh} L^m_I L^\ah_J L^\bh_K)$ is the 
Wess-Zumino term which is constructed such that ${\cal S}_{GS}$
is invariant under $\kappa$-symmetry.
Under $G\to\Omega G H$ for global $\Omega$ and local $H$, 
the currents $G^{-1}\p G$
are invariant up to a tangent-space Lorentz rotation using the
standard coset construction where 
$[P_m,Q_\a,Q_{\overline\a}, J_{mn}]$ are the generators in $\Omega$ and
$J_{mn}$ are the generators in $H$.
So as long as the action is constructed from
Lorentz-invariant combinations of currents, the action is
invariant under the global target-space isometries generated by
$[P_m,Q_\a,Q_{\overline\a}, J_{mn}]$.
Note that because the R-R field-strength is self-dual,
only $(3D-10)$ of the $\half D(D-1)$ Lorentz generators $J_{mn}$ appear
in \commrel. So only $(3D-10)$ of the $L^{mn}$ currents are nonzero in
\current.

The terms $d_\a \overline L^\a$ and $\overline d_\ah L^\ah$ in \altact\ break
kappa symmetry but allow quantization since they imply non-vanishing
propagators for $\t^\a$ and $\overline\t^\ah$. And the term 
$-\half
d_\a \overline d_\bh F^{\a\bh}$ comes from the R-R vertex operator and implies
that certain components of $d_\a$ and $\overline d_\bh$ are auxiliary fields.
For the $D=10$ background, the term ${\cal S}_{comp}$ is absent, while
for the $D=6$ and $D=4$ backgrounds, ${\cal S}_{comp}$ is the action
for an $N=2$ $c={3\over 2}(10-D)$ superconformal field theory
which describes the $(10-D)$-dimensional compactification manifold.

Finally, ${\cal S}_{ghost}$ describes the action for the worldsheet ghosts.
This action
is non-trivial in the $D=10$ background since the $D=10$ ghosts transform
under Lorentz transformations and therefore couple through their
Lorentz currents to the spacetime
connection and curvature of the background. In $D=10$, this ghost action
is 
\eqn\ghostact{{\cal S}_{ghost}= 
\int d^2 z [{\cal L}_{ghost}^{flat} + \half N_{mn} \overline L^{mn} 
+ \half\overline N_{mn} L^{mn} +{1\over 4} N_{mn}\overline N_{pq}
R^{mnpq}]}
where ${\cal L}_{ghost}^{flat}$ is the free Lagrangian in a flat background
for the left and right-moving worldsheet ghosts $(\l^\a,w_\a)$
and $(\overline\l^\ah,\overline w_\ah)$, $\l^\a$ and $\overline\l^\ah$ are
pure spinors satisfying $\l\g^m\l=\overline\l\g^m\overline\l=0$ for $m=0$ to 9,
$w_\a$ and $\overline w_\ah$ are their conjugate momenta,
$N_{mn}= \half \l\g_{mn}w$ and 
$\overline N_{mn}= \half \overline\l\g_{mn}
\overline w$ are their left and right-moving
Lorentz currents, and $R^{mnpq}$ is the target-space curvature tensor.
Note that ${\cal S}_{ghost}$ is invariant under local tangent-space
Lorentz rotations, which is necessary for the action to be well-defined
on the coset superspace described by $G(x,\t,\overline\t)$.
In the $D=6$ and
$D=4$ actions, the 
worldsheet ghosts are Lorentz scalars so their Lorentz currents
vanish and ${\cal S}_{ghost} ={\cal S}_{ghost}^{flat}$.

The various terms in \altact\ have been chosen such that the $D=10$
action is BRST invariant and such that the $D=6$ and $D=4$
actions are N=2 worldsheet superconformally invariant.
To check these invariances at the classical level, it is useful to
compute the equations of motion for $d_\a$ and $\overline d_\ah$.

Suppose one varies $Z^M=[x^m,\t^\a,\overline\t^\ah]$ such that 
$E^\a_M \d Z^M=\rho^\a$,
$E^\ah_M \d Z^M= \overline\rho^\ah$, and $E^m_M \d Z^M=0$ where
$L^\a= E^\a_M \p Z^M$,
$L^\ah= E^\ah_M \p Z^M$, $L^m = E^m_M \p Z^M$,
and $[L^\a,L^\ah,L^m]$ are defined in \current. Then the covariant
GS action ${\cal S}_{GS}$ transforms as 
\eqn\transgs{\d {\cal S}_{GS}= 2\rho^\a L^m \g_{m\a\b}\overline L^\b +
2\overline\rho^\ah \overline L^m \g_{m\ah\bh} L^\bh.} The transformation of
\transgs\ is related to kappa symmetry since when $\rho^\a = 
\kappa_\b L^m \g_m^{\a\b}$ and
$\overline\rho^\ah = 
\overline\kappa_\bh L^m \g_m^{\ah\bh}$, $\d {\cal S}_{GS}$ is proportional
to the Virasoro constraints $\eta_{mn} L^m L^n$ and 
$\eta_{mn}\overline L^m \overline L^n$.

Furthermore, the commutation relations of \commrel\ imply that
\eqn\transcur{\d L^\a =\p\rho^\a + {1\over 4}
(\g^{mn})^\a_\b L_{mn} \rho^\b
+F^{\a\bh}\g^m_{\bh\gh} L_m  \overline\rho^\gh,}
$$\d L^\ah =\p\overline\rho^\ah + {1\over 4}
(\g^{mn})^\ah_\bh L_{mn} \overline\rho^\bh
-F^{\b\ah}\g^m_{\b\g} L_m \rho^\g,$$
$$\d L^{mn} = (\g^{[m} F \g^{n]})_{\b\gh}\rho^\b L^\gh 
+(\g^{[m} F \g^{n]})_{\b\gh} L^\b \overline\rho^\gh$$
where 
$(\g^{[m} F \g^{n]})_{\a\dh}=\half( \g^m_{\a\b} F^{\b\gh} \g^n_{\gh\dh}-
\g^n_{\a\b} F^{\b\gh} \g^m_{\gh\dh}).$

So by varying $\rho^\a$ and $\overline\rho^\ah$, one obtains the equations
of motion
\eqn\motiond{\overline\p d_\a= 2\g^m_{\a\b} L_m \overline L^\b + {1\over 4}
d_\b (\g_{mn})_\a^\b
\overline L^{mn} - \overline d_\bh F^{\g\bh} \g^m_{\g\a} L_m
 +\half  
(\g^{[m} F \g^{n]})_{\a\gh}(N_{mn} \overline L^\gh +\overline N_{mn} L^\gh),}
$$\p\overline d_\ah=2 \g^m_{\ah\bh} \overline L_m L^\bh + {1\over 4}
\overline d_\bh (\g_{mn})_\ah^\bh
L^{mn} +  d_\b F^{\b\gh} \g^m_{\gh\ah} \overline L_m
 -\half  
(\g^{[m} F \g^{n]})_{\g\ah}(N_{mn} \overline L^\g +\overline N_{mn} L^\g).$$

Plugging into \motiond\ the equations of motion $\overline L^\a=\half F^{\a\bh}
\overline d_\bh$ and 
$L^\ah= -\half F^{\b\ah}
d_\b$ which come from varying $d_\a$ and $\overline d_\ah$, one finds 
\eqn\motd{\overline\nabla d_\a= 
\half(\g^{[m} F \g^{n]})_{\a\gh}(N_{mn} \overline L^\gh -\half
\overline N_{mn} F^{\d\gh} d_\d),}
$$\nabla\overline d_\ah=
-\half(\g^{[m} F \g^{n]})_{\g\ah}
(\half N_{mn} F^{\g\overline\d} \overline d_{\overline\d}
+\overline N_{mn} L^\g),$$
where the spin connections in the covariantized derivatives
$\nabla$ and $\overline\nabla$ are $L^{mn}$
and $\overline L^{mn}$.

When $D=10$, BRST invariance implies that the left and
right-moving BRST operators, $\l^\a d_\a$
and $\overline\l^\ah \overline d_\ah$, must be holomorphic and antiholomorphic.
To check that this is implied by \motd, note that the equations of motion
of $\l^\a$ and $\overline \l^\ah$ coming from \ghostact\ are\chandia 
\eqn\motlambda{
\overline\nabla 
\l^\a= {1\over 8} R^{mnpq} (\g_{mn})_\b^\a \l^\b \overline N_{pq},}
$$\nabla \overline\l^\ah = {1\over 8}
R^{mnpq} (\g_{pq})_\bh^\ah \overline\l^\bh N_{mn}.$$
So \motd\ and \motlambda, together with the identity
of \curvid,
imply that
\eqn\motq{\overline\p(\l^\a d_\a)= \half\l^\a 
(\g^{[m} F \g^{n]})_{\a\gh} N_{mn}\overline L^\gh,}
$$\p(\overline\l^\ah \overline d_\ah)= -\half\overline\l^\ah 
(\g^{[m} F \g^{n]})_{\g\ah} N_{mn} L^\g.$$
Since $N_{mn}=\half (\l\g_{mn} w)$ and $\l^\a\l^\b$ is proportional
to $(\l\g^{pqrst}\l) (\g_{pqrst})^{\a\b}$, the right-hand side of \motq\
is proportional to 
$\g_{mn}\g_{pqrst}\g^{[m} F \g^{n]} $. But since
$\g_m \g_{pqrst} \g^m=0$, one finds that  
\eqn\propzero{\g_{mn}\g_{pqrst}\g^{[m} F \g^{n]}  =
\g_{pqrst}\g^n F \g_n  =\g_{pqrst} \g^n \g_{uvwxy} \g_n F^{uvwxy}=0.}
So $\overline\p(\l^\a d_\a)= \p(\overline\l^\ah \overline d_\ah)=0$ as desired.

When $D=4$ and $D=6$, N=2 worldsheet superconformal invariance implies that
the left and right-moving superconformal generators must be holomorphic
and antiholomorphic. In these actions, 
\motd\ implies that
$\overline\nabla 
d_\a=\nabla \overline d_\ah=0$ since $N_{mn}$ and $\overline N_{mn}$ vanish.
And since the left and right-moving 
N=2 superconformal constraints in the $D=4$ and
$D=6$ formalisms
are constructed out of Lorentz-invariant combinations of 
$d_\a$ and $\overline d_\ah$, \motd\ implies that these constraints are
holomorphic and antiholomorphic.

Up to now, the analysis of the action for the Ramond-Ramond
plane wave background has been identical to the analysis of
the action for the $AdS_{D\over 2}\times S^{D\over 2}$ background. The
only difference between the backgrounds is that the $(2D-4)\times(2D-4)$
matrix $F^{\a\bh}$ is invertible for the 
$AdS_{D\over 2}\times S^{D\over 2}$
background, whereas $F^{\a\bh}$ is
not invertible and has rank $D-2$ for the R-R plane wave background.
However,
as will now be shown, this difference
considerably simplifies the quantum analysis of the action in the R-R
plane wave background.
Although the $AdS_{D\over 2}\times S^{D\over 2}$ action has only been
proven to be conformally invariant at the one-loop level\bz\foot{
For the $AdS_5\times S^5$ action, one-loop conformal invariance has not
yet been proven for the ghost contribution \ghostact\
to the action. For the ``harmonic'' version of the $AdS_3\times S^3$
action in \adswitten, exact conformal invariance has been proven.}, it will 
be possible to prove exact conformal invariance for the action in 
a Ramond-Ramond plane wave background.

\newsec{Conformal Invariance of the Action in an R-R 
Plane Wave Background}

In a Ramond-Ramond plane wave background, the only nonzero components
of the self-dual field strength are $F^{- j_1 ... j_{(D-2)\over 2}}$
where $j$ ranges over the light-cone directions $j=1$ to $(D-2)$. It
is convenient to split the $SO(D-1,1)$ spinor representation labeled
by $\a$ and $\ah$ into $SO(D-2)$ representations labeled by
$(a,a')$ 
and $(\overline a,\overline a')$ where $(a,a',\overline a,\overline a')$
range from 1 to $(D-2)$. Using this notation,
\eqn\lcnot{ (\g^-)_{a b}=\d_{a b},\quad
(\g^-)_{a' b'}=0,\quad
(\g^+)_{a b}=0,\quad
(\g^+)_{a' b'}=\d_{a' b'},}
$$(\g^-)_{\overline a \overline b}=\d_{\overline a \overline b},\quad
(\g^-)_{\overline a' \overline b'}=0,\quad
(\g^+)_{\overline a \overline b}=0,\quad
(\g^+)_{\overline a' \overline b'}=\d_{\overline a' \overline b'}.$$

Since $F^{a \overline b'}
=F^{a' \overline b'}= F^{a' \overline b}=0$, the commutation
relations of \commrel\ imply that
\eqn\planecom{[P^-,P^j]= \mu^2 J^{+j},\quad
\{Q_a,Q_b\}=2 P^+\d_{ab},\quad
\{Q_{\overline a},Q_{\overline b}\}=2 P^+\d_{\overline a\overline b},}
$$[P^-,Q_a]= - \d_{ab} F^{b\overline c}Q_{\overline c},\quad
[P^-,Q_{\overline a}]= \d_{\overline a\overline b} F^{c\overline b}Q_{c},$$
where $\mu^2= F^{a\overline b} 
F^{c\overline d}\d_{ac}\d_{\overline b\overline d}$.
Therefore, $[Q_{a'}, Q_{\overline a'}, J_{jk}]$
are never created from commutators of $[P_m,Q_a,Q_{\overline a}]$. 
So 
$[L^{a'},L^{\overline a'}, L^{jk}]$ in \current\ only depend on
$(\t^{a'},\overline\t^{\overline a'})$ and are independent of
$(x^m,\t^a,\overline\t^{\overline a})$. 
This implies 
that if $\t^{a'}$ and 
$\overline\t^{\overline a'}$ are defined to carry charge $+1$
and $d_{a'}$ and $\overline d_{\overline a'}$ are defined to carry charge $-1$,
all terms in the action of 
\altact\ carry non-negative charge. 

The term with zero charge in \altact\ is 
\eqn\altz{{\cal S}_{(0)}=
{\cal S}_{GS}{}|_{\t^{a'}=\overline\t^{\overline a'}=0}
+\int d^2 z (d_a \overline L^a_{(0)} + d_{a'} \overline\p\t^{a'} + 
\overline d_{\overline a} 
L^{\overline a}_{(0)} 
+ \overline d_{\overline a'} \p\overline\t^{\overline a'} -
{1\over 2}
d_a \overline d_{\overline b} F^{a\overline b}) }
$$+ S_{comp} 
+ \int d^2 z[{\cal L}_{ghost}^{flat} +N_{-j} \overline L^{-j}_{(0)}
+ \overline N_{-j} L^{-j}_{(0)}
+ N_{-j} N_{-k} \eta^{jk}]$$ 
where $[L^m_{(0)},L^a_{(0)},L^{\overline a}_{(0)},L^{-j}_{(0)}]$=
$[L^m,L^a,L^{\overline a},L^{-j}]|_{\t^{a'}=\overline\t^{\overline a'}=0}.$
Also, at $\t^{a'}=\overline\t^{\overline a'}=0$, one can use the commutation
relations of \commrel\ to show that the Wess-Zumino term in
\covgs\ simplifies to $\half(\overline L^a_{(0)} L^{\overline b}_{(0)}-
L^a_{(0)} \overline L^{\overline b}_{(0)}) F^{-1}_{a\overline b}$ where 
$F^{-1}_{a\overline b}$ is the inverse to the $(D-2)
\times(D-2)$ matrix $F^{a \overline b}$.
So one can write
\eqn\covgsex{{\cal S}_{GS}{}|_{\t^{a'}=\overline\t^{\overline a'}=0}=
\int d^2 z [\half\eta_{mn} 
L^m_{(0)} L^n_{(0)} + \half F^{-1}_{a\overline b} (\overline
L^a_{(0)} L_{(0)}^{\overline b}-
L_{(0)}^a  \overline L_{(0)}^{\overline b})].}
After
integrating out $d_a$ and $\overline d_{\overline a}$, one obtains the action
\eqn\szero{{\cal S}_{(0)}=
\int d^2 z [\half \eta_{mn}L^m \overline L^n - 
\half F^{-1}_{a\overline b} (3 \overline L_{(0)}^a L_{(0)}^{\overline b}+
L_{(0)}^a \overline L_{(0)}^{\overline b})}
$$
+ d_{a'}\overline\p\t^{a'}
+ \overline d_{\overline a'}\p\overline\t^{\overline a'} +
{\cal L}_{ghost}^{flat}
+ 
N_{-j} \overline L^{-j}_{(0)}
+ \overline N_{-j} L^{-j}_{(0)} + N_{-j} N_{-k} \eta^{jk}]+{\cal S}_{comp}.$$ 

As will now be shown, ${\cal S}_{(0)}$ 
is conformally invariant. This can be used to prove
conformal invariance of the action of \altact\ since all
terms with positive charge in \altact\ are related to $S_{(0)}$ by
isometries of the action. In other words, 
the global isometries of the background imply that the action is constructed
from the currents of \current\
in combinations which are invariant under tangent-space Lorentz
transformations.
These combinations are 
\eqn\combic{\eta_{mn}L^m\overline L^n,\quad 
\int dy \e^{IJK} (\g_{m\a\b}L^m_I L^\a_J L^\b_K +
\g_{m\ah\bh} L^m_I L^\ah_J L^\bh_K), \quad d_\a \overline L^\a,\quad
\overline d_{\ah} L^{\ah},}
$$d_\a \overline d_{\bh} F^{\a\bh}, \quad N_{mn} \overline N_{pq} R^{mnpq},
\quad 
{\cal L}_{ghost}^{flat} + 
\half(N_{mn}\overline L^{mn} +\overline N_{mn} L^{mn}),$$
and only the coefficients 
in front of the various combinations
can be adjusted without breaking the isometries. However, the coefficients are
determined once one knows $S_{(0)}$, so if $S_{(0)}$ is conformally invariant,
the entire action of \altact\ is conformally invariant. 

To show that $S_{(0)}$ is conformally invariant, note that the first line
of \szero\ has precisely the $G/H$ coset space structure discussed in \bz\
where ${\cal H}_0= J^{+j}$, 
${\cal H}_1= Q_{a}$, 
${\cal H}_2= P^m$, 
and ${\cal H}_3= Q_{\overline a}$. Using the analysis 
of \bz, one can therefore prove that the first line of \szero\ is one-loop
conformally invariant. Furthermore, one can prove the exact conformal 
invariance of ${\cal S}_{(0)}$ by computing the currents
\eqn\currtwo{G^{-1} \p G= P_m L^m_{(0)} + Q_a L^a_{(0)} +
Q_{\overline a} L^{\overline a}_{(0)} + J_{-j} L^{-j}_{(0)}}
where $G(x^m,\t^a,\overline\t^{\overline a})= \exp (x^+ P^-)\exp (x^- P^+ +
x^j P^j + \t^a Q_a +\overline\t^{\overline a}Q_{\overline a})$. One finds that
\eqn\ssimple{{\cal S}_{(0)}=
\int d^2 z [\half \p x^m\overline\p x_m -
2 F^{-1}_{a\overline b}\p\t^a\overline\p\overline\t^{\overline b}
+\half \p x^+\overline \p x^+ \mu^2 x^j x^j}
$$
+\p x^+ \d_{ab}\t^a 
\overline \p\t^b + \overline\p x^+ \d_{\overline a\overline b}
\overline\t^{\overline a}\p\overline\t^{\overline b}
+ d_{a'}\overline\p \t^{a'} 
+\overline d_{\overline a'} \p\overline\t^{\overline a'}
+ {\cal L}_{ghost}^{flat}$$
$$
+  \mu^2(x^j (\l^{b'}\s_j^{a c'}\d_{b' c'} w_a \overline\p x^+ +
\overline\l^{\overline b'}\s_j^{\overline a \overline c'}
\d_{\overline b'\overline c'} \overline w_{\overline a} 
\p x^+ )
+
(\l^{b'}\s_j^{a c'}\d_{b' c'} w_a )
(\overline\l^{\overline b'}
\s^{j\overline a \overline c'}
\d_{\overline b'\overline c'} \overline w_{\overline a}))] + 
{\cal S}_{comp}.$$

By separating the worldsheet variables in \ssimple\ into background
values and quantum variables, and integrating over the quantum variables,
one computes the quantum effective action. Since $[x^-,\l^a,w_{a'},
\overline\l^{\overline a},
\overline w_{\overline a'}]$ appear in the quadratic part of
\ssimple\ but do not appear in the vertices, the variables
$[x^+,w_a,\l^{a'},
\overline w_{\overline a},
\overline \l^{\overline a'}]$ can be set to their background
value in the vertices of \ssimple. After doing this, all quantum variables
appear at most quadratically in \ssimple, which means they can only give
a one-loop contribution to the effective action. This one-loop contribution
is easily computed
to vanish where the $x^+$ dependence cancels after integrating
over the $x^j$ and $[\t^a,\overline\t^{\overline a}]$ quantum variables, and
the central charge cancels for the same reason as in a flat background.

It has therefore been proven that the action of \altact\ for the superstring
in an R-R plane wave background is an exact conformal field theory.
It would be interesting to try to use this conformal field theory to
compute scattering amplitudes. Since this conformally invariant action
does not require interaction point operators, the amplitude computations
might be simpler than using the light-cone gauge action. Although the
action of \altact\ is more complicated than the quadratic
light-cone action, there might be certain amplitude computations in
which ``charge conservation'' of the $(\t^{a'},\overline\t^{\overline a'})$
variables implies
that the complicated action of \altact\ can be replaced by the simpler
action of \ssimple.

{\bf Acknowledgements:} 

NB would like to thank Chris Hull, Juan
Maldacena, Arkady Tseytlin, and especially Michael Green
for useful
discussions, the Newton Institute for their
hospitality, and CNPq grant 300256/94-9, Pronex grant
66.2002/1998-9 and FAPESP grant
99/12763-0 for partial financial support.
This research was partially conducted during the period that NB
was employed by the Clay Mathematics Institute as a CMI Prize Fellow. 

\listrefs
\end